\documentstyle[11pt,epsfig]{article}
\title{\bf Spherically symmetric solutions and gravitational
 collapse in brane-worlds}
\author{Malihe Heydari-Fard $^{1}$\thanks{email:
 m.heydarifard@mail.sbu.ac.ir}
and Hamid R. Sepangi $^{1}$\thanks{email: hr-sepangi@sbu.ac.ir}
\\ {\small $^{1}$Department
of Physics, Shahid Beheshti University, Evin, Tehran 19839, Iran}}
\begin{document}
\maketitle %\baselineskip 24pt
\begin{abstract}
We consider spherically symmetric solutions within the context of
brane-world theory without mirror symmetry or any form of junction
conditions. For a constant curvature bulk, we obtain the modified
Tolman-Oppenheimer-Volkoff (TOV) interior solutions in two cases
where one is matched to a schwarzschild-de Sitter exterior while
the other is consistent with an exterior solution whose structure
can be used to explain the galaxy rotation curves without
postulating dark matter. We also find the upper bound to the mass
of a static brane-world star and show that the influence of the
bulk effects on the interior solutions is small. Finally, we
investigate the gravitational collapse on the brane and show that
the exterior of a collapsing star can be static in this scenario.
\vspace{5mm}\\
PACS numbers: 04.50.-h, 04.20.Jb, 11.25.Mj, 04.70.Bw
\end{abstract}
\section{Introduction}
The idea that our familiar 4-dimensional ($4D$) space-time is a
hypersurface (brane) in a 5-dimensional space-time (bulk)
\cite{Nima,Randall,Dvali} has been under detailed elaboration
during the last decade. According to this brane-world scenario,
all matter and gauge interactions reside on the brane, while
gravity can propagate in the whole 5-dimensional space-time.
Several brane-world cosmologies have been proposed in the context
of Randall-Sundrum (RS) formulations \cite{Randall}, defined in a
5-dimensional anti-de Sitter space-time ($AdS_5$). The dynamics of
these models feature boundary terms in the action and sometimes
mirror symmetry, such that bulk gravitational waves interfere with
the brane-world motion. This usually comes together with junction
conditions producing an algebraic relationship between the
extrinsic curvature and the confined matter \cite{israel,Battye}.
The consequence is that the Friedman equation acquires an
additional term which is proportional to the square of energy
density of the confined matter field \cite{Cline,Binetruy}. This
term was initially considered as a possible solution to the
accelerated expansion of the universe. However, soon it was
realized to be incompatible with the big bang nucleosynthesis,
requiring additional fixes \cite{Binetruy}.

Brane-world scenarios under more general conditions and still
compatible with the brane-world program have also been rather
extensively studied over the past decade where it has been shown
that it is possible to find a richer set of cosmological solutions
in accordance with the current observations \cite{Maia}. Under
these conditions, without using $Z_2$ symmetry or without
postulating any junction condition, Friedman equation is modified
by a geometrical term which is defined in terms of the extrinsic
curvature, leading to a geometrical interpretation for dark energy
\cite{maia}. There have also been arguments concerning the
uniqueness of the junction conditions. Indeed, other forms of
junction conditions exist, so that different conditions may lead
to different physical results \cite{Battye}. Furthermore, these
conditions cannot be used when more than one non-compact extra
dimension is involved. Against this background, an interesting
higher-dimensional model was introduced in \cite{Rubakov} where
particles are trapped on a 4-dimensional hypersurface by the
action of a confining potential. The dynamics of test particles
confined to a brane by the action of such potential at the
classical and quantum levels were studied in \cite{shahram}. In
\cite{fard}, a brane-world model was studied in which matter is
confined to the brane through the action of such a potential
without using any junction conditions, offering a geometrical
explanation for the accelerated expansion of the universe. A
geometrical explanation for the generalized Chaplygin gas was
considered in \cite{gas} along the same line. We have also studied
exact solutions of the vacuum field equations on the brane for two
interesting cases. The first solution can be used to explain the
galaxy rotation curves without assuming the existence of dark
matter and without having to resort to the Modified Newtonian
Dynamics (MOND), and the second solution represents a black hole
in an asymptotically de Sitter space-time \cite{Razmi}.

One physically important problem in  brane-world scenarios is the
development of a full understanding of stellar structures and
black holes. Static, spherically symmetric exterior vacuum
solutions of the brane-world models were first proposed by Dadhich
and co-workers \cite{naresh} and Germani and Maartens
\cite{maartens}. In \cite{naresh}, the authors obtained an exact
black hole solution of the effective Einstein equation on the
brane under the condition that the bulk has non-zero Weyl
curvature and the brane spacetime satisfies the null energy
condition. The solution is given by the usual Reissner-Nordstrom
(RN) metric where the charge parameter is thought of as a tidal
charge arising from the projection of the Weyl curvature of the
bulk onto the brane. The tidal charge, like the RN electric
charge, would generate a $1/r^2$ term in the potential while the
high energy modification to the Newtonian potential cannot be any
stronger than $1/r^3$ \cite{Randall,Tanaka}. The cause for this
disagreement is the presence of tidal charges which is a measure
of the bulk Weyl curvature. The main drawback of the solution is
that we do not know the corresponding bulk solution. It is however
agreed that the RN metric is a good approximation to a black hole
on the brane near the horizon \cite{Reall}. It has also been shown
that the vacuum field equations on the brane reduce to a system of
two ordinary differential equations which describe all the
geometrical properties of the vacuum as functions of dark pressure
and dark radiation terms \cite{mak}. Stellar structure in
brane-world models is very different from that in ordinary general
relativity. An exact interior uniform density stellar solution on
the brane has been found in \cite{maartens}. In this model the
general relativistic upper bound for the mass-radius ratio,
$M<\frac{4}{9}R$, is reduced by 5-dimensional high-energy effects
\cite{maartens}. Spherically symmetric brane-world solutions when
there is a contribution from the brane intrinsic curvature
invariant in the dynamics have been studied in
\cite{Kofinas,Kofinas1}.

The gravitational collapse on the brane has been widely studied by
many authors \cite{Collapse}-\cite{collapse}. Based on the tidal
charge scenario, Oppenheimer-Snyder type \cite{snyder}
gravitational collapse of spherically symmetric objects was
analyzed in \cite{Collapse}. This was formulated by a no-go
theorem that indicates a non-static exterior for the collapsing
sphere on the brane. The non-static exterior of the collapsing
brane star could be the Vaidya radiating solution on the brane
\cite{dad}. The non-static nature of a collapsing brane star for
induced gravity with or without the Gauss-Bonnet term have also
been studied in \cite{kofinas}. However, it was demonstrated in
\cite{gergely} that a static exterior can be obtained by relaxing
the idea of dust inside the star, thereby introducing a
non-vanishing surface pressure, and by ignoring the tidal effect.
It has also been shown that a generalized non-empty bulk may lead
to a manifestly static exterior for a collapsing spherical star on
the brane \cite{pal}.

In this paper, following the model introduced in \cite{Maia,maia},
we consider a 4-dimensional brane embedded in a 5-dimensional
bulk, without using the $Z_2$ symmetry or without postulating any
junction condition. Taking a constant curvature bulk, the
effective field equations on the brane are modified by an extra
term, $Q_{\mu\nu}$, which is a geometrical quantity. We study the
interior space-time of stars in this scenario and derive the
modified Tolman-Oppenheimer-Volkoff (TOV) equations on the brane
in two cases. We investigate gravitational collapse of spherical
objects on the brane and show the possibility of having a static
exterior for a collapsing sphere in this scenario.
\section{Field equations and matching conditions}
The embedding of the brane-world in the bulk plays an essential
role on the covariant formulation of the brane-world gravity,
since it tells us how the Einstein-Hilbert dynamics of the bulk is
transferred to the brane-world. However, there are many different
ways to embed a manifold into another, classified as local,
global, isometric, conformal, rigid, deformable, analytic or
differentiable. The choice of one or other depends on what the
embedded manifold is supposed to do.

Generally, there are three basic postulates in the geometrical
approach considered in brane-world scenarios, that is, the
confinement of the standard gauge interactions to the brane, the
existence of quantum gravity in the bulk and finally, the
embedding of the brane-world. All other model dependent properties
such as warped metric, mirror symmetries, radion or extra scalar
fields, fine tuning parameters like the tension of the brane and
the choice of a junction condition are left out as much as
possible in our calculations \cite{Maia}.

In the following we present a brief review of the model proposed
in \cite{maia,maiaa}. Consider a 4-dimensional brane
$(\Sigma,g_{\mu\nu})$ embedded in a $m$-dimensional bulk $(M,{\cal
G}_{AB})$. The components of the Riemann tensor of the bulk
written in the embedding vielbein $\{{\cal Z}^{A}_{, \alpha},
{\cal N}^A_a \}$, lead to the Gauss-Codazzi equations,
respectively \footnote{Capita Latin indices refer to the bulk
dimensions. Small case Latin indices refer to the extra dimensions
and all Greek indices refer to the brane.} \cite{Book}
\begin{eqnarray}\label{a1}
R_{\alpha \beta \gamma \delta}=2g^{ab}K_{\alpha[ \gamma
a}K_{\delta] \beta b}+{\cal R}_{ABCD}{\cal Z} ^{A}_{,\alpha}{\cal
Z} ^{B}_{,\beta}{\cal Z} ^{C}_{,\gamma} {\cal Z}^{D}_{,\delta},
\end{eqnarray}
\begin{eqnarray}\label{a2}
D_{\delta}K_{\alpha \gamma c}-D_{\gamma}K_{\alpha\delta
c}=2g^{ab}A_{[\gamma ac}K_{ \delta] \alpha b}+{\cal R}_{ABCD}{\cal
Z} ^{A}_{,\alpha} {\cal N}^{B}_{c} {\cal Z} ^{C}_{,\gamma} {\cal
Z}^{D}_{,\delta},
\end{eqnarray}
where $D_{\mu}$ is the covariant differentiation with respect to
$g_{\mu\nu}$. Also, ${\cal N}_{a}^{A}$ $(a,b=4,...,m)$ are the
components of the $(m-4)$-independent normal vectors to $\Sigma$
and the induced metric on $\Sigma$ is $g^{\mu\nu}{\cal
Z}^{A}_{,\mu}{\cal Z}^{B}_{,\nu}={\cal G}^{AB}-g^{ab}{\cal
N}^{A}_{a}{\cal N}^{B}_{b}$. Contracting the Gauss equation
(\ref{a1}) on ${\alpha}$ and ${\gamma}$ we find
\begin{eqnarray}\label{a3}
R_{\mu\nu}=(K_{\mu\alpha c}K_{\nu}^{\,\,\,\,\alpha c}-K_{c} K_{\mu
\nu }^{\,\,\,\ c})+{\cal R}_{AB} {\cal Z}^{A}_{,\mu} {\cal
Z}^{B}_{,\nu}-g^{ab}{\cal R}_{ABCD}{\cal N}^{A}_{a}{\cal
Z}^{B}_{,\mu}{\cal Z}^{C}_{,\nu}{\cal N}^{D}_{b}.
\end{eqnarray}
A further contraction gives the Ricci scalar
\begin{eqnarray}\label{a4}
R=(K_{\mu\nu a}K^{\mu\nu a}-K_{a} K^{a})+{\cal R}-2g^{ab}{\cal
R}_{AB}{\cal N}^{A}_{a}{\cal N}^{B}_{b}+g^{ad}g^{bc}{\cal
R}_{ABCD}{\cal N}^{A}_{a}{\cal N}^{B}_{b}{\cal N}^{C}_{c}{\cal
N}^{D}_{d}.
\end{eqnarray}
Therefore the Einstein-Hilbert action for the bulk geometry in
$m$-dimensions can be written as \cite{maia}
\begin{eqnarray}\label{action}
\int({\cal R}-2\Lambda^{(b)})\sqrt{{\cal G}}d^mx
&\equiv&\int\left[R-2\Lambda^{(b)}-(K_{\mu\nu a}K^{\mu\nu a}-K_{a}
K^{a})\right]\sqrt{{\cal
G}}d^mx\nonumber\\
&+& \int\left[2g^{ab}{\cal R}_{AB}{\cal N}^{A}_{a}{\cal N
}^{B}_{b}-g^{ad}g^{bc}{\cal R}_{ABCD}{\cal N}^{A}_{a}{\cal
N}^{B}_{b}{\cal N}^{C}_{c}{\cal N}^{D}_{d}\right]\sqrt{{\cal
G}}d^mx.
\end{eqnarray}
The covariant equations of motion for a brane-world in a
$m$-dimensional bulk can be derived by taking the variation of
(\ref{action}) with respect to $g_{\mu\nu}$ and $g_{ab}$, noting
that the Lagrangian depends on these variables through ${\cal
Z}_{,\mu}^{A}$ \cite{maia}. Thus, the field equation for
$g_{\mu\nu}$, with the confined matter represented by
$\tau_{\mu\nu}$ is
\begin{eqnarray}\label{a5}
R_{\mu\nu}-\frac{1}{2}Rg_{\mu\nu}=\alpha^{*}\tau_{\mu\nu}-\Lambda^{(b)}g_{\mu\nu}+Q_{\mu\nu}+S_{\mu\nu},
\end{eqnarray}
where we have denoted
\begin{eqnarray}\label{a6}
Q_{\mu\nu}=g^{ab}\left(K^{\rho}_{\mu a}K_{\rho\nu b}-K_aK_{\mu\nu
b}\right)-\frac{1}{2} \left(K_{\alpha\beta a}K^{\alpha\beta
a}-K_aK^a\right)g_{\mu\nu},
\end{eqnarray}
and
\begin{eqnarray}\label{a7}
S_{\mu\nu}=g^{ab}{\cal R}_{AB}{\cal N}^{A}_{a}{\cal
N}^{B}_{b}g_{\mu\nu}-g^{ad}{\cal R}_{ABCD}{\cal N}^{A}_{a}{\cal
Z}^{B}_{,\mu}{\cal Z}^{C}_{,\nu}{\cal N}^{D}_{d}.
\end{eqnarray}
The last term $S_{\mu\nu}$ in equation (\ref{a5}) depends on the
definition of the geometry of the bulk \cite{maiaa}. Now, we
restrict our analysis to a 5-dimensional bulk with a constant
curvature characterized by the Riemann tensor
\begin{eqnarray}\label{2}
{\cal R}_{ABCD}=k_{*}({\cal G}_{AC}{\cal G}_{BD}-{\cal
G}_{AD}{\cal G}_{BC}),\label{2}
\end{eqnarray}
where $k_*$ denotes the bulk constant curvature. In the flat case
$k_*=0$ and in the de Sitter and anti-de Sitter cases we may write
$k_*=\pm\frac{\Lambda^{(b)}}{6}$ respectively. In the normal
Gaussian frame defined by the embedded space-time the bulk metric
may be decomposed as
\begin{eqnarray}
{\cal G}_{AB}=\left( \!\!\!
\begin{array}{cc}
g_{\mu \nu } & 0 \\
0 & g_{55}
\end{array}
\!\!\!\right) ,\hspace{.5 cm}g_{55}=+1,
\end{eqnarray}
with this assumptions the Gauss-Codazzi equations reduce to
\begin{eqnarray}
R_{\alpha\beta\gamma\delta} =
(K_{\alpha\gamma}K_{\beta\delta}-K_{\alpha\delta}K_{\beta\gamma})
+ k_{*}
(g_{\alpha\gamma}g_{\beta\delta}-g_{\alpha\delta}g_{\beta\gamma}),\label{3}
\end{eqnarray}
\begin{eqnarray}
K_{\alpha[\beta;\gamma]} = 0,\label{4}
\end{eqnarray}
and $S_{\mu\nu}=3k_{*}g_{\mu\nu}$. Thus the dynamical equation
(\ref{a5}) given by
\begin{eqnarray}
G_{\mu\nu}=\alpha^{*}\tau_{\mu\nu}- \lambda g_{\mu\nu}+
Q_{\mu\nu},\label{6}
\end{eqnarray}
where
\begin{eqnarray}
Q_{\mu\nu}=\left(K^{\rho}_{\,\,\,\,\mu
}K_{\rho\nu}-KK_{\mu\nu}\right)-\frac{1}{2}
\left(K_{\alpha\beta}K^{\alpha\beta}-K^2\right)g_{\mu\nu}.\label{7}
\end{eqnarray}
Here, $\tau_{\mu\nu}$ is the confined matter energy-momentum
tensor on the brane and $\lambda=-3k_{*}+\Lambda^{(b)}$. Using
equation (\ref{4}) and considering the definition of $Q_{\mu\nu}$,
we find
\begin{eqnarray}
Q^{\mu\nu}_{\,\,\,\,;\mu}=0.\label{8}
\end{eqnarray}
Thus, $Q_{\mu\nu}$ is independently a conserved quantity so that
there is no exchange of energy between this geometrical correction
and the confined matter. Such an aspect has one important
consequence, that is, if $Q_{\mu\nu}$ is to be related to dark
energy, it does not exchange energy with ordinary matter, much the
same as in coupled quintessence models \cite{W}. The confined
matter source on the brane is considered to be an isotropic
perfect fluid
\begin{eqnarray}
\tau_{\mu\nu}=(\rho+p )u_{\mu}u_{\nu}+p g_{\mu\nu},\hspace{.5
cm}p=(\gamma-1)\rho,
\end{eqnarray}
where $\rho=p=0$ for the exterior solutions ($r>R$).

In order to solve the Codazzi equation (\ref{4}), we choose the
static spherically symmetric metric on the brane in the form
\begin{eqnarray}
ds^2=-e^{\mu(r)}dt^2+e^{\nu(r)}dr^2+r^2\left(d\theta^2+\sin^2\theta
d\varphi^2\right).\label{9}
\end{eqnarray}
The York relation
\begin{eqnarray}
K_{\mu \nu }=-\frac{1}{2}\frac{\partial
g_{\mu\nu}}{\partial\xi},\label{10}
\end{eqnarray}
then shows that in a diagonal metric, $K_{\mu\nu }$ are diagonal.
Now, separating the spatial components, the Codazzi equation
(\ref{4}) reduces to
\begin{equation}
K_{\mu\nu ,\sigma}-K_{\nu\rho }\Gamma^{\rho}_{\mu\sigma}=
K_{\mu\sigma ,\nu}-K_{\sigma\rho
}\Gamma^{\rho}_{\mu\nu},\label{12}
\end{equation}
\begin{eqnarray}
K_{00,1}-\left(\frac{\mu^{'}}{2}\right)K_{00}=-\left(\frac{\mu^{'}e^\mu}{2e^\nu}\right)
K_{11},\label{13}
\end{eqnarray}
\begin{eqnarray}
K_{22,1}-\left(\frac{1}{r}\right)K_{22}=\left({re^{-\nu}}\right)
K_{11},\label{14}
\end{eqnarray}
where a prime represents differentiation with respect to $r$. The
first equation gives
$K_{00,\sigma}=K_{11,\sigma}=K_{22,\sigma}=K_{33,\sigma}=0$ for
$\sigma=0,3$. Repeating the same procedure for $\sigma=2$, we
obtain $K_{00,\sigma}=K_{11,\sigma}=K_{22,\sigma}=0$. This shows
that $K_{11}$ depends only on the variable $r$ and the choice
$K_{11}=\alpha e^{\nu(r)}$ would simplify our analysis. Using
equations (\ref{13}),(\ref{14}) and $K_{11}$, one finds
\begin{eqnarray}
K_{00}(r)=-\alpha e^{\mu(r)}+ce^{\mu(r)/2},\label{15}
\end{eqnarray}
\begin{eqnarray}
K_{22}(r)=\alpha r^2+\beta r.\label{16}
\end{eqnarray}
Taking $\mu,\nu=3$ in the first equation we obtain
\begin{eqnarray}
K_{33,1}-\left(\frac{1}{r}\right)K_{33}=\left({e^{-\nu}r\sin^2{\theta}}
\right)K_{11}=\alpha r\sin^2{\theta},\label{17}
\end{eqnarray}
\begin{eqnarray}
K_{33,2}-\left(\cot{\theta}\right)K_{33}=\left
(\sin{\theta}\cos{\theta}\right)K_{22}.\label{18}
\end{eqnarray}
Using equations (\ref{15}), (\ref{16}) and (\ref{17}), we find
\begin{eqnarray}
K_{33}(r,\theta)=\alpha r^2\sin^2{\theta}+r\beta
\sin^2{\theta}+rc_{1}\sin{\theta}.\label{19}
\end{eqnarray}
Now, use of equation (\ref{7}) leads to the components of
$Q_{\mu\nu}$
\begin{eqnarray}\label{EQ}
Q_{00}&=&-\frac{g_{00}}{ r^2} \left[3\alpha^2r^2+4\alpha\beta
r+\beta^2+\frac{c_1}{\sin{\theta}}(2\alpha
r+\beta) \right],\nonumber\\
Q_{11}&=&- \frac{g_{11}}{ r^2}\left[3\alpha^2r^2+4\alpha\beta
r+\beta^2+\frac{c_1}{\sin{\theta}}\left(2\alpha r+\beta-c
re^{-\mu/2}\right)-2ce^{-\mu/2}\left(\alpha r^2+\beta
r\right)\right], \nonumber\\
Q_{22}&=& \frac{g_{22}}{
r}\left[-3\alpha^2r-2\alpha\beta+ce^{-\mu/2}(2\alpha
r+\beta)+\frac{c_1}{\sin{\theta}}
\left(-2\alpha+ce^{-\mu/2}\right)\right],\nonumber\\
Q_{33}&=& \frac{g_{33}}{
r}\left[-3\alpha^2r-2\alpha\beta+ce^{-\mu/2}(2\alpha
r+\beta)\right].
\end{eqnarray}
Since $G_{2}^{2}=G_{3}^{3}$ and thus $Q_{2}^{2}=Q_{3}^{3}$, one
obtains $c_1=0$. Now, using these relations and equation
(\ref{6}), the gravitational field equations become
\begin{eqnarray}\label{eq1}
{e^{-\nu}}\left(-\frac{1}{r^2}+\frac{\nu^{'}}{r}\right)+\frac{1}{r^2}=
\alpha^{*}\rho+\lambda+3\alpha^2+\frac{4\alpha\beta}
{r}+\frac{\beta^2}{r^2},
\end{eqnarray}
\begin{eqnarray}\label{eq2}
e^{-\nu}\left(\frac{1}{r^2}+\frac{\mu^{'}}{r}\right)-\frac{1}{r^2}=
\alpha^{*}p-\lambda-3\alpha^2-\frac{4\alpha\beta}
{r}-\frac{\beta^2}{r^2}+2ce^{-\mu/2}\left(\alpha+\frac{\beta}{r}\right),
\end{eqnarray}
\begin{eqnarray}\label{eq3}
e^{-\nu}\left(\frac{\mu^{'}-\nu^{'}}{2r}-\frac{\mu^{'}\nu^{'}}{4}
+\frac{\mu^{''}}{2}+\frac{\mu^{'2}}{4}\right)=
\alpha^{*}p-\lambda-3\alpha^2-\frac{2\alpha\beta}{r}+2ce^{-\mu/2}\left(\alpha+\frac{\beta}{2r}\right).
\end{eqnarray}
We again note that the exterior solutions are characterized by
$\rho=p=0$.

Using the contracted Bianchi identities and equation (\ref{8}) the
conservation equations is given by
\begin{eqnarray}
\tau^{\mu\nu}_{\,\,\,\,\,;\nu}=0.
\end{eqnarray}
For the static, spherical symmetry metric (\ref{9}), these
equations give
\begin{eqnarray}\label{pressure}
p^{'}+\frac{\mu^{'}}{2}\left(\rho+p\right)=0.
\end{eqnarray}
The Israel-Darmois matching conditions at the stellar surface
$\Sigma$ are given by \cite{Israel}
\begin{eqnarray}\label{22}
[G_{\mu\nu}r^{\nu}]_{\Sigma}=0,
\end{eqnarray}
where $[f]_{\Sigma}\equiv f(R^{+})-f(R^{-})$. Using equations
(\ref{6}) and (\ref{22}) and taking $\lambda=0$, one finds
\begin{eqnarray}\label{23}
[\alpha^{*}\tau_{\mu\nu}r^{\nu}+Q_{\mu\nu}r^{\nu}]_{\Sigma}=0,
\end{eqnarray}
which, upon using the second equation in (\ref{EQ}), leads to
\begin{eqnarray}\label{24}
p(R)=0.
\end{eqnarray}
In our model this result coincides with that of general relativity
whereas in brane-world models where a delta-function in the
energy-momentum is used, we take this as an assumption
\cite{maartens}. In the next section, we study the influence of
$Q_{\mu\nu}$ term on the interior brane-world solutions ($r\leq
R$).
\section{Exact solutions with uniform-density}
Equations (\ref{eq1})-(\ref{eq3}) and (\ref{pressure}) determine
the system of field equations on the brane. In what follows we
will consider the cases where there are three independent field
equations which imply energy-momentum conservation, namely, $c=0$
and $c=\beta=0$. One may then either use the three independent
field equations or two of the field equations together with the
energy-momentum conservation equation. For the purpose of this
paper it is more convenient to follow the latter approach. Now,
taking $c=0$ and $\lambda=0$, the field equations become
\begin{eqnarray}\label{29}
{e^{-\nu}}\left(-\frac{1}{r^2}+\frac{\nu^{'}}{r}\right)+\frac{1}{r^2}=
\alpha^{*}\rho+3\alpha^2+\frac{4\alpha\beta}
{r}+\frac{\beta^2}{r^2},
\end{eqnarray}
\begin{eqnarray}\label{30}
e^{-\nu}\left(\frac{1}{r^2}+\frac{\mu^{'}}{r}\right)-\frac{1}{r^2}=
\alpha^{*}p-3\alpha^2-\frac{4\alpha\beta} {r}-\frac{\beta^2}{r^2},
\end{eqnarray}
\begin{eqnarray}\label{31}
p^{'}+\frac{\mu^{'}}{2}\left(\rho+p\right)=0.
\end{eqnarray}
Equation (\ref{29}) can easily be integrated to give
\begin{eqnarray}\label{32}
e^{\nu(r)}=\left[1-\frac{m(r)}{r}\right]^{-1},
\end{eqnarray}
where the mass function is
\begin{eqnarray}\label{33}
m(r)=\alpha^{*}\int^{r}_{s}\left[\rho(r^{'})+\frac{3\alpha^2}{\alpha^{*}}+\frac{4\alpha\beta}
{\alpha^{*}r^{'}}+\frac{\beta^2}{\alpha^{*}r^{'2}}\right]r^{'2}dr^{'},
\end{eqnarray}
and $s=0$, $s=R$ represent the interior and exterior solutions
respectively. If an equation of state of the form $\rho=\rho(p)$
is assumed for the interior, the conservation equation (\ref{31})
can be integrated to give
\begin{eqnarray}\label{34}
e^{\mu(r)}=\exp\left[-2\int^{p(r)}_{p_{c}}\frac{d
p}{p+\rho(p)}\right],
\end{eqnarray}
where $p_{c}$ denotes the central pressure. The function
${\mu^{'}}$ can be eliminated from equations (\ref{eq2}) and
(\ref{pressure}), yielding the modified TOV equation
\begin{eqnarray}\label{35}
\frac{p^{'}}{p+\rho}=-\frac{1}{2}\frac{\left[\alpha^{*}pr+
\frac{m(r)}{r^2}-3\alpha^2r-4\alpha\beta-\frac{\beta^2}{r}+2c r
e^{-\mu/2}(\alpha r+\beta)\right]}{\left[1-\frac{m(r)}{r}\right]}.
\end{eqnarray}
In the general relativity limit of
$\alpha,\beta,c\longrightarrow0$, we regain the usual TOV
equation. Taking the simplest case of a static spherical star with
uniform density given by $\rho(r)=\rho_0=\frac{3M}{4\pi R^3}$, we
can find the exact interior solutions,  generalizing the interior
solutions predicted in general relativity. Stars with such a
uniform distribution are of interest, not only because they
actually exist, but also because they are simple enough to allow
an exact solution of the Einstein field equations and set an upper
limit to the gravitational red shift of spectral lines from the
surface of any star \cite{weinberg}.

Integration of equation (\ref{33}) immediately gives
\begin{eqnarray}\label{36}
e^{\nu(r)}=\frac{1}{1-\frac{\alpha^{*}M}{4\pi
r}(\frac{r}{R})^{3}-(\alpha r+\beta)^2},\hspace{.5 cm}r\leq R.
\end{eqnarray}
Furthermore, integration of equation (\ref{31}) results in
\begin{eqnarray}\label{37}
e^{\mu(r)}=\frac{B}{\left(\rho_{0}+p(r)\right)^2},\hspace{.5
cm}r\leq R,
\end{eqnarray}
where $B$ is a constant of integration. Also the pressure is given
by
\begin{eqnarray}\label{38}
p(r)=-\rho_{0}\frac{be^{\frac{\nu(r)}{2}}\left[\alpha\beta b_1
r^3+\left(6\alpha^2\beta^2b_3-\frac{b_1b_2}{3}\right)r^2-5\alpha\beta
\left(\frac{18\alpha^2b_3}{5\alpha^{*}\rho_0}+b_2b_3\right)r+b_2b_3\left(1-\beta^2\right)\right]+
I(r)}{I(r)\left[1+be^{\frac{\nu(r)}{2}}\left(\alpha\beta
r+\beta^2-1\right)\right]},
\end{eqnarray}
where
\begin{eqnarray}\label{388}
I(r)=b_1r^2+6b_3\alpha\beta r+3b_3(\beta^2-1),
\end{eqnarray}
and
\begin{eqnarray}\label{3888}
b_1&=&\alpha^{*}\rho_0+6\alpha^2+\frac{9\alpha^4}{\alpha^{*}\rho_0},
\nonumber\\b_2&=&1-\beta^2-\frac{6\alpha^2}{\alpha^*\rho_0},\nonumber\\b_3&=&1+\frac{3\alpha^2}{\alpha^{*}\rho_0}.
\end{eqnarray}
In the above equation, $b$ is the constant of integration and is
evaluated by taking $p(R)=0$. The vanishing of the pressure at the
surface, which is certainly physically reasonable, is a
consequence of the application of the Israel matching conditions
at the stellar surface. In the limit
$\alpha,\beta\longrightarrow0$, equation (\ref{38}) reduces to the
pressure for a uniform-density star in general relativity.

The exterior solutions of equations (\ref{eq1})-(\ref{eq3}) are
found to be \cite{Razmi}
\begin{eqnarray}\label{39}
e^{-\nu(r)}=1-\frac{A_1}{r}-\alpha^2r^2-2\alpha\beta r-\beta^2,
\end{eqnarray}
and
\begin{eqnarray}\label{399}
e^{\mu(r)}=\frac{f(r)}{4r}\left(-A_2+2\alpha c
\int\frac{r^{5/2}dr}{f(r)^{3/2}}+2\beta
c\int\frac{r^{3/2}dr}{f(r)^{3/2}}\right)^{2},
\end{eqnarray}
where $A_1$ and $A_2$ are integration constants and
\begin{eqnarray}\label{3999}
f(r)=-r+A_1+\alpha^2r^3+2\alpha\beta r^2+\beta^2r.
\end{eqnarray}
There are a number of arbitrary constants which do not let us find
the unique vacuum solution of the gravitational field equations on
the brane because the Birkhoff theorem does not apply here
\cite{maartens,mak}.

Equation (\ref{399}) cannot be solved in closed form. The choice
$c=0$ leads to the exact exterior solution
\begin{eqnarray}\label{40}
e^{\mu(r)}=e^{-\nu(r)}=1-\frac{A_1}{r}-\alpha^2r^2-2\alpha\beta
r-\beta^2,\hspace{.5 cm}r\geq R.
\end{eqnarray}
This solution can be used to explain the galactic rotation curves
without relying on the existence of dark matter and without
assuming any new modified theory, {\it e.g.} MOND. The matching of
the interior solution to that of the exterior then determines the
integration constant as $A_1=2GM$.

A second class of solutions of the system of equations
(\ref{eq1})-(\ref{eq3}) can be obtained by the choice $c=\beta=0$
and $\alpha\neq0$. Substituting $c=\beta=0$ into equation
(\ref{EQ}), one obtains
\begin{eqnarray}\label{Q}
Q_{00}=-3\alpha^2{g_{00}},\hspace{.5 cm}
Q_{\mu\nu}=-3\alpha^2g_{\mu\nu},\hspace{.5 cm}\mu,\nu=1,2,3.
\end{eqnarray}
As we noted before, $Q_{\mu\nu}$ is an independently conserved
quantity, suggesting an analogy with the energy-momentum tensor of
an uncoupled non-conventional energy source. Let us define
$Q_{\mu\nu}$ as an isotropic perfect fluid and write
\begin{eqnarray}\label{25}
Q_{\mu\nu}\equiv\frac{1}{\alpha^*}\left[(\rho_{extr}+p_{extr})
u_{\mu}u_{\nu}+p_{extr} g_{\mu\nu}\right],
\end{eqnarray}
where we have denoted the ``geometric pressure'' associated with
the extrinsic curvature by $p_{extr}$ and the ``geometric energy
density'' by $\rho_{extr}$. The geometric fluid can be implemented
by the equation of state
\begin{eqnarray}
p_{extr}=(\gamma_{extr}-1)\rho_{extr},\label{extr}
\end{eqnarray}
where $\gamma_{extr}$ may be a function of the radius. Comparing
$Q_{\mu\nu}$ and $Q_{00}$ from equation (\ref{25}) with the
components of $Q_{\mu\nu}$ given by equation (\ref{Q}), we obtain
\begin{eqnarray}
p_{extr}=-\frac{3\alpha^2}{\alpha^*}, \hspace{.5 cm}
\rho_{extr}=\frac{3\alpha^2}{\alpha^*}.
\end{eqnarray}
Equation (\ref{extr}) then gives $\gamma_{extr}=0$, showing that
the geometrical matter may play the role of a positive
cosmological constant
$(3\alpha^2=\Lambda\approx3\times10^{-56}cm^2)$ on the interior
space-time \cite{C}. In this case, substituting $\beta=0$ into
equations (\ref{36}) and (\ref{37}), the interior line element
takes the form
\begin{eqnarray}\label{44}
ds^2=-\frac{B}{\left(\rho_{0}+p(r)\right)^2}dt^2+\frac{dr^2}
{1-(\frac{\alpha^{*}\rho_0}{3}+\alpha^2)r^2}+r^2(d\theta^2+\sin^2\theta
d\varphi^2).
\end{eqnarray}
Since $\frac{\alpha^{*}\rho_0}{3}+\alpha^2>0$ the spatial geometry
of this metric describes part of a 3-sphere of radius
$\frac{1}{\sqrt{{\alpha^{*}\rho_0/3}+\alpha^2}}$ and has a
coordinate singularity at
$r=\hat{r}=\frac{1}{\sqrt{{\alpha^{*}\rho_0/3}+\alpha^2}}$. The
metric is well defined for radii less than $\hat{r}$. The pressure
is given by
\begin{eqnarray}\label{41}
p(r)=\rho_{0}\frac{
\left(1-\frac{6\alpha^{2}}{\alpha^{*}\rho_0}\right)-p_{0}\sqrt{1-(\frac{\alpha^{*}\rho_0}{3}+\alpha^2)r^2}}
{-3+p_{0}\sqrt{1-(\frac{\alpha^{*}\rho_0}{3}+\alpha^2)r^2}},
\end{eqnarray}
where the integration constant $p_{0}$, evaluated by defining
$p(r=0)=p_c$ to be the central pressure, is given by
\begin{eqnarray}\label{p0}
p_{0}=\frac{3p_c+\rho_0-\frac{6\alpha^2}{\alpha^{*}}}{p_c+\rho_0}.
\end{eqnarray}
For coordinate singularity $\hat{r}$ the pressure is
\begin{eqnarray}
p(\hat{r})=p_{s}\equiv-\frac{\rho_0}{3}\left(1-\frac{6\alpha^2}{\alpha^*
\rho_0}\right),
\end{eqnarray}
which is negative if $(1-\frac{6\alpha^2}{\alpha^{*}\rho_0})>0$.
Since $p(\hat{r})<0$ there exists an $R$ such that $p(R)=0$,
consistent with equation (\ref{24}). Taking $p(R)=0$, $R$ can be
derived from equation (\ref{41}) as
\begin{eqnarray}\label{pc}
\sqrt{1-\left(\frac{\alpha^{*}\rho_0}{3}+\alpha^2\right)R^2}=
\frac{1}{p_{0}}\left(1-\frac{6\alpha^{2}}{\alpha^{*}\rho_0}\right).\label{pc}
\end{eqnarray}
Now, using equations (\ref{p0}) and (\ref{pc}) the central
pressure is
\begin{eqnarray}
p_{c}=\rho_{0}\frac{\left(1-\frac{6\alpha^2}{\alpha^*
\rho_0}\right)\left[1-\sqrt{1-\left(\frac{\alpha^{*}\rho_0}{3}+\alpha^2\right)R^2}\right]}
{3\sqrt{1-\left(\frac{\alpha^{*}\rho_0}{3}+\alpha^2\right)R^2}-\left(1-\frac{6\alpha^2}{\alpha^*
\rho_0}\right)},
\end{eqnarray}
and
\begin{eqnarray}
p(r)=\rho_{0}\frac{
\left(1-\frac{6\alpha^{2}}{\alpha^{*}\rho_0}\right)-
\left(1-\frac{6\alpha^{2}}{\alpha^{*}\rho_0}\right)\left(\frac{e^{\nu(R)}}{e^{\nu(r)}}\right)^
{\frac{1}{2}}}
{-3+\left(1-\frac{6\alpha^{2}}{\alpha^{*}\rho_{0}}\right)
\left(\frac{e^{\nu(R)}}{e^{\nu(r)}}\right)^{\frac{1}{2}} }.
\end{eqnarray}
We can also obtain an upper limit on compactness from the
requirement that $p(r)$ must be finite. Figure 1 shows the
behavior of pressure as a function of $r$.
\begin{figure}
\centerline{\begin{tabular}{ccc}
\epsfig{figure=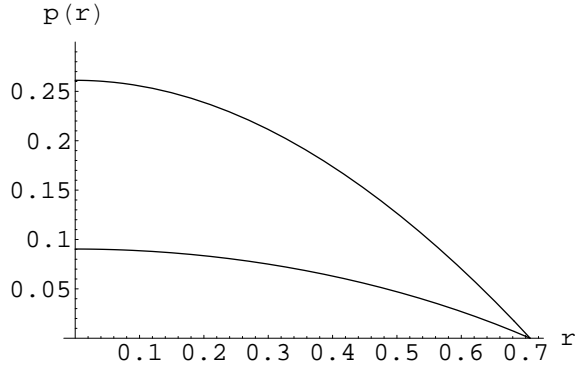,width=8cm}
\end{tabular} } \caption{\footnotesize Pressure as a function of radius in general relativity
(bottom curve) and in the brane-world model with
$(1-\frac{6\alpha^2}{\alpha^*\rho_0})>0$ (top curve).}\label{fig1}
\end{figure}
As can be seen the pressure is a decreasing function of $r$,
similar to what one obtains in general relativity. This is
equivalent to the conditions that $p_c$ is finite and positive
which gives the following condition for the mass
\begin{eqnarray}\label{new}
MG<\frac{4}{9}R-\frac{2}{9}R\left(1-\sqrt{1-\frac{9}{4}\alpha^2R^2}\right).
\end{eqnarray}
For a given $R$ there is an upper bound to the mass of a static
star where the central pressure becomes infinite as $M \rightarrow
M_{max}$. For the choice $3\alpha^2=\Lambda$, the correction to
the general relativity limit of $\frac{4}{9}$ is small in this
scenario. The square root term is real if
\begin{eqnarray}
R\leq\frac{2}{3\alpha}.
\end{eqnarray}
Also, definition of $p_s$ above implies
$\alpha^2<\frac{\alpha^{*}\rho_{0}}{6}$. Definition of mass can
then be used to re-write this to give
\begin{eqnarray}
3\alpha^2R^2<\frac{\alpha^*\rho_{0}R^2}{2}<\frac{2}{3}+\sqrt{\frac{4}{9}-\alpha^2R^2},
\end{eqnarray}
which reduces to
\begin{eqnarray}
R<\frac{1}{\sqrt{3}\alpha},
\end{eqnarray}
indicating that the boundary of the stellar object is located
before the event horizon is reached. Also, if we take $\alpha=0$
we recover the standard general relativity result, namely
\begin{eqnarray}
\frac{MG}{R}<\frac{4}{9}.
\end{eqnarray}
For the case $c=\beta=0$ and $\alpha\neq0$ the exterior solution
is given by
\begin{eqnarray}\label{42}
e^{\mu(r)}=e^{-\nu(r)}=1-\frac{A_1}{r}-\alpha^2r^2,\hspace{.5
cm}r\geq R,
\end{eqnarray}
where $A_1$ is an integration constant. The matching condition
implies $A_1=2GM$ and therefore
\begin{eqnarray}\label{43}
B=\rho_{0}^2\left(1-\frac{2GM}{R}-\alpha^2R^2\right).
\end{eqnarray}
The corresponding line element now takes the form
\begin{eqnarray}\label{M}
ds^2=-\left(1-\frac{2GM}{r}-\alpha^2r^2\right)dt^2+\frac{dr^2}
{\left(1-\frac{2GM}{r}-\alpha^2r^2\right)}
+r^2\left(d\theta^2+\sin\theta^2d\varphi^2\right).\label{45}
\end{eqnarray}
Comparing the above result with the line element for the black
hole solution in an asymptotically de Sitter space, the
cosmological constant is found to be $3\alpha^2=\Lambda$. This
positive value is in agreement with present observations. Note
that for $\sqrt{27}GM<\frac{1}{\alpha}$ there are two horizons
$2GM<r_1<6GM$ and $\frac{1}{\sqrt{3}\alpha}<r_2<\frac{1}{\alpha}$,
while for $\sqrt{27}GM=\frac{1}{\alpha}$, $r_1$ and $r_2$ coincide
and there is only one horizon $r=\frac{1}{\sqrt{3}\alpha}$. Note
also that there is no horizon for $\sqrt{27}GM>\frac{1}{\alpha}$.

In the special case where
$1-\frac{6\alpha^2}{\alpha^{*}\rho_0}=0$, the line element takes
the form
\begin{eqnarray}
ds^2=-\frac{B}{\left(\rho_{0}+p(r)\right)^2}dt^2+\frac{dr^2}
{(1-3\alpha^2r^2)}+r^2(d\theta^2+\sin^2\theta d\varphi^2),
\end{eqnarray}
and the pressure is given by
\begin{eqnarray}
p(r)=\rho_{0}\frac{p_{0}\sqrt{1-3\alpha^2r^2}}{3-p_{0}\sqrt{1-3\alpha^2r^2}}.
\end{eqnarray}
We note that the pressure vanishes at the coordinate singularity
which means that the stellar surface coincides with the coordinate
singularity. Thus the interior solution can not be joined to the
schwarzschild-de Sitter exterior solution.
\section{Gravitational collapse on the brane}
In this section, we discuss the gravitational collapse for
spherically symmetric objects in our model. For a sphere
undergoing Oppenheimer-Snyder collapse, the collapsing region can
be conveniently expressed by a Robertson-Walker metric
\begin{eqnarray}\label{C1}
ds^2=-d\tau^2+a(\tau)^2(1+\frac{k\chi^2}{4})^{-2}\left[d\chi^2+\chi^2\left(d\theta^2+\sin^2\theta
d\varphi^2\right)\right].
\end{eqnarray}
The modified Friedman equation from equation (\ref{6}) is
\cite{maia}
\begin{eqnarray}\label{C2}
\frac{\dot{a}^2}{a^2}=\frac{\alpha^{*}\rho_0}{3}a^{-3}
+b_0^2a^{-3\gamma_{extr}}+\frac{\lambda}{3}-\frac{k}{a^2}.
\end{eqnarray}
Also, the Raychaudhuri equation can be written as
\begin{eqnarray}
\frac{\ddot{a}}{a}=-\frac{\alpha^{*}\rho_0}{6}a^{-3}
+b_0^2a^{-3\gamma_{extr}}\left(1-\frac{3}{2}\gamma_{extr}\right)+\frac{\lambda}{6},
\end{eqnarray}
where $\gamma_{extr}$ is defined by equation (\ref{extr}) and
$b_0$ is an integration constant. The modified Friedman equation
can also be written in terms of the proper radius, $r(\tau)={\chi
a(\tau)}/{(1+\frac{k\chi^2}{4})}$, of the collapsing boundary
surface at $\chi=\chi_{0}$, that is
\begin{eqnarray}\label{C3}
{\dot{r}}^2=\frac{\alpha^{*}\rho_0\chi_0^{3}}{3}\left(1+\frac{k\chi_{0}^2}{4}\right)^{-3}
\frac{1}{r}+b_0^{2}\chi_{0}^{3\gamma_{extr}}
\left(1+\frac{k\chi_{0}^2}{4}\right)^{-3\gamma_{extr}}r^{-3\gamma_{extr}+2}+\frac{\lambda}{3}r^2+E,
\end{eqnarray}
where
\begin{eqnarray}
E=-\frac{k\chi_0^{2}}{\left(1+\frac{k\chi_0^2}{4}\right)^2},
\end{eqnarray}
is the energy per unit physical mass. Let us express the static
spherically symmetric metric for the vacuum exterior by
\begin{eqnarray}\label{C4}
ds^2=-F(r)^2e^{\mu(r)}dt^2+e^{-\mu(r)}dr^2
+r^2\left(d\theta^2+\sin^2\theta d\varphi^2\right).
\end{eqnarray}
In order for a metric of the form (\ref{C4}) to become the
exterior to the region described by metric (\ref{C1}), the metric
and extrinsic curvature have to be continuous across the
collapsing boundary surface. The two conditions are simultaneously
satisfied by expressing the metrics (\ref{C1}) and (\ref{C4}) in
terms of null coordinates by adopting the method of
\cite{Collapse}, which results in $F(r)=1$ and
\begin{eqnarray}\label{C9}
e^{\mu(r)}=1-\frac{\alpha\rho_0\chi_0^{3}}{3}\left(1+\frac{k\chi_{0}^2}{4}\right)^{-3}
\frac{1}{r}-b_0^{2}\chi_{0}^{3\gamma_{extr}}
\left(1+\frac{k\chi_{0}^2}{4}\right)^{-3\gamma_{extr}}r^{-3\gamma_{extr}+2}.
\end{eqnarray}
The choice $\gamma_{extr}=0$, $b_0^2=\alpha^2$ and
$\frac{\alpha\rho_0\chi_0^{3}}{3}(1+\frac{k\chi_0^4}{4})^{-3}=A_1$,
then leads to
\begin{eqnarray}\label{C10}
e^{\mu(r)}=1-\frac{A_1}{r}-\alpha^2r^2.
\end{eqnarray}
Equations (\ref{C4}) and (\ref{C10}) imply the brane Ricci scalar
is
\begin{eqnarray}\label{Ricci}
R^{\mu}_{\,\,\,\,\mu}=12\alpha^2.
\end{eqnarray}
Equation (\ref{6}) for a vacuum exterior implies
\begin{eqnarray}\label{Ricci1}
R^{\mu}_{\,\,\,\,\mu}=-Q^{\mu}_{\,\,\,\,\mu}=12\alpha^2+\frac{12\alpha\beta}{r}+
\frac{2\beta^2}{r^2}-2ce^{-\mu/2}\left(3\alpha+\frac{2\beta}{r}\right).
\end{eqnarray}
Comparing equations (\ref{Ricci}) and (\ref{Ricci1}) we see that a
static exterior is possible only if $c=\beta=0$. It is now obvious
that the static exterior solution (\ref{C9}), surrounding the
collapsing region, can take the form of equation (\ref{42}).
Considering the universe as a 3-dimensional brane embedded in a
5-dimensional bulk of constant curvature, without $Z_2$ symmetry
or any form of junction conditions, the vacuum exterior of a
spherical cloud can be static [27,28], which is similar to the
standard general relativity and is different from the brane-world
models where a delta-function is used \cite{Collapse} to confine
matter on the brane.
\section{Conclusions}
In this paper, we have studied spherically symmetric solutions in
a brane-world model without mirror symmetry or any form of
junction conditions. We have shown that within the context of the
model presented here, the matching conditions lead to a vanishing
pressure at the surface of the star. This result is different from
those obtained in Randall-Sundrum type brane-world models where
the vanishing of pressure at the surface of the star is simply
assumed. We have obtained exact uniform-density stellar solutions
localized on a 3-brane in two cases by considering a constant
curvature bulk. The first solution is consistent with an exterior
solution whose structure can be used to explain the galaxy
rotation curves and the second solution represents a stellar model
with the exterior Schwarzschild-de Sitter space-time. We have also
obtained the upper bound to the mass of a static brane-world star,
for the case $c=\beta=0$, and shown that the influence of the bulk
on the interior solutions is small. Finally, we have studied the
fate of a collapsing star on the brane and shown that the exterior
of a collapsing star can be static in our model.

\end{document}